## Limb Spicules from the Ground and from Space

### Jay M. Pasachoff • William A. Jacobson • Alphonse C. Sterling

**Abstract**  We amassed statistics for quiet-sun chromosphere spicules at the limb using ground-based observations from the *Swedish 1-m Solar Telescope* on La Palma and simultaneously from NASA's *Transition Region and Coronal Explorer* (*TRACE*) spacecraft.  The observations were obtained in July 2006.  With the 0.2 arcsecond resolution obtained after maximizing the ground-based resolution with the Multi-Object Multi-Frame Blind Deconvolution (MOMFBD) program, we obtained specific statistics for sizes and motions of over two dozen individual spicules, based on movies compiled at 50-second cadence for the series of five wavelengths observed in a very narrow band at Hα, on-band and at ±0.035 nm and ±0.070 nm (10 s at each wavelength) using the SOUP filter, and had simultaneous observations in the 160 nm EUV continuum from *TRACE*.  The MOMFBD restoration also automatically aligned the images, facilitating the making of Dopplergrams at each off-band pair.  We studied 40 Hα spicules, and 14 EUV spicules that overlapped Hα spicules; we found that their dynamical and morphological properties fit into the framework of several previous studies. From a preliminary comparison with spicule theories, our observations are consistent with a reconnection mechanism for spicule generation, and with UV spicules being a sheath region surrounding the Hα spicules.

J. M. Pasachoff
Williams College—Hopkins Observatory, Williamstown, MA 01267, USA, and
California Institute of Technology 150-21, Pasadena, CA 91125, USA
email: jay.m.pasachoff@williams.edu

W. A. Jacobson
Williams College—Hopkins Observatory, Williamstown, MA 01267, USA

A. C. Sterling, NASA/MSFC, VP62/Space Science Office, Huntsville, AL 35805
Current address: JAXA/Institute of Space and Astronautical Science, Hinode Group, 3-1-1
Yoshinodai, Sagamihara, Kanagawa 229-8510, Japan

# 1. Introduction

The solar chromosphere is made almost entirely or entirely (certainly for the upper chromosphere) of jets known as spicules, with lifetimes of approximately 15 minutes with hundreds of thousands on the sun at any given time (Beckers, 1968). There have been numerous investigations over the decades of spicules, for example, Roberts (1945), Lippincott (1957), and Dunn (1960) studied them at the polar limb. A spectroscopic study at the Sacramento Peak Observatory by Pasachoff, Noyes, and Beckers (1968) provided direct evidence of spicular motion, overcoming limitations of apparent motion derived from series of intensity images that may have resulted merely from changing ionization fronts instead of actual velocities. In addition to rising and falling motion along the limb, Pasachoff, Noyes, and Beckers also reported evidence for rotation of spicules. Mouradian (1965, 1967) also provided spectroscopic measurements of spicular motions, and Lynch, Beckers, and Dunn (1973) and Nishikawa (1988) later remeasured size statistics. See also Zaqarashvili and Erdélyi (2009).

Generally, the term spicules refers to the features seen at the limb of the quiet Sun (Golub and Pasachoff, 2001), and this paper deals exclusively with such "classical" spicules. Similar features seen on the disk are called "mottles," although some authors refer to those features as spicules as well. Beckers (1968, 1972) gives comprehensive reviews of earlier work on spicules, as well as mottles. Active consideration of limb spicules revived at the time of the 1998 total eclipse with a meeting on *Solar Jets and Coronal Plumes* on Guadaloupe (Koutchmy, Martens, and Shibata, 1998). Reviews of spicules were given by Suematsu (1998) and, from theoretical considerations, by Sterling (1998a). Those presenting new observations included Salakhutdinov and Papushev (1998), Zirin and Cameron (1998), and De Pontieu et al. (1998), Budnik et al. (1998), and Dara, Koutchmy, and Suematsu (1998).

Sterling's paper included numerical simulations based on the deposition of thermal energy in the middle or upper chromosphere, perhaps as microflares (Sterling, Shibata, and Mariska, 1994; Sterling et al., 1991; Sterling, 1998b).

Observations in the UV and EUV also reveal features with spicule-like properties. For example, from observations made from a rocket, Dere, Bartoe, and Brueckner (1983) observed EUV chromospheric jets that they identified with spicules though their lifetimes were about 10 times shorter. Chae et al. (1999) have used TRACE data together with their own Big Bear Solar Observatory data to comment on EUV jets and their relation to solar microflares. They dwelt on the Fe XII images at 195 Å from TRACE, comparing them with Hα in an active region on the disk. They did not study limb spicules.

One reason for the voluminous literature on spicules is because they are difficult to observe from the ground, since their widths are very small, 1 arcsec or less. This sub-groundbased-resolution status has made it difficult to specify with confidence their properties. In turn, this has made it difficult to differentiate among differing spicule models. Most such models are based on waves or shocks generated by a physical or thermal impulse near the base of a magnetic flux tube, as reviewed by Sterling (2000). In that review, Sterling concluded that most or all of the theories as of that time had difficulties, seemingly being inconsistent with one or more aspects of the observation-based picture of spicules at that time. The observations themselves however, did not yield a consistent coherent picture of spicules with which to compare the models.  A new era of spicule observations and theory was needed.

There has been substantial progress in recent years through new high-resolution ground-based observations, and with observations in EUV, from the *TRACE* satellite.  Combining observations of on-disk mottles seen in absorption in EUV with results from numerical models, De Pontieu, Erdélyi, and James (2004) found that photospheric p-mode oscillations could explain at least one variety of spicule-like features.  (De Pontieu and Haerendel (1998) and De Pontieu (1999) were among those with earlier models of spicules; they evaluated spicules as weakly-damped Alfvén waves.)  While traveling through the chromosphere, waves formed by the oscillations propel chromospheric material into the corona, and that material has properties closely matching the observed features.  Subsequent studies show that the same mechanism can act on dynamic fibrils and Hα mottles also (e.g., Hansteen et al. 2006, Rouppe van de Noort et al., 2007).

Meanwhile, other high-resolution ground-based observations appear to reveal a new feature on the disk, features that are finer than the traditional mottles.  Rutten (2006, 2007) observed these features using the Dutch Open Telescope, and termed them "straws.''  Even more recently, observations with the Solar Optical Telescope (SOT; Tsuneta et al. 2008) on the *Hinode* satellite (Kosugi et al. 2007), shows these features in more detail, especially at the solar limb in Ca II images.  De Pontieu et al. (2007b) refer to the straw-like features as "Type II spicules," while reserving the term "Type I spicules" for the above-mentioned features formed by p-mode waves.  The Type II spicules (also discussed by, e.g., Langangen et al. 2008) are reported to be a few 100 km thick, to have apparent motions of $50 - 150$ km/s, lifetimes of $\sim 10 - 60$ s, and to have a different origin from that of the Type I spicules.

One objective of the current paper is to solidify the observational properties of spicules at the solar limb, as observed from the ground.  Just what the relationship is between the recently named Type I and Type II spicules (both types are seen both on the disk and at the limb; examples of both at the limb are in De Pontieu et al. 2007a, Fig. 3) and the spicules traditionally observed at the chromospheric limb in Hα, is a deep mystery.  Clearly, over the near-term future, observations from *Hinode* will be key to understanding spicules.  In this work we set the background for that currently ongoing and future work with *Hinode*, by measuring spicule properties with the currently best-available ground-based instrumentation.  Our results here, when eventually combined with future work with *Hinode* and other data, will help to untangle the connections between the "new-era spicules" being seen now, with Hinode on the one hand, and the "traditional spicules" that have been observed, measured, and argued about for generations with ground-based instrumentation on the other hand.

Another unclear point about traditional spicules is their relationship to the spicule-like features seen in UV and EUV (see, e.g., Xia et al. 2005).  In particular, it is not certain whether they are some aspect of Hα spicules, or if they are totally independent features (see discussion and references in Sterling 2000).  Some recent work has investigated on-disk UV and Hα features (e.g., de Wijn & De Pontieu 2006), also comparing *TRACE* and *Swedish 1-m Solar Telescope* imaging, showing that at least some chromospheric features are related at the differing wavelengths.  Here we will continue investigations into this question by exploring properties of UV spicules seen with the *TRACE* satellite.

Our study of limb spicules presented here takes advantage of the exceptional resolution sometimes obtainable with the *Swedish 1-m Solar Telescope* (*SST*) of the Royal Swedish Academy of Sciences, located on La Palma in the Canary Islands, Spain. The highest seeing quality requires both favorable atmospheric conditions and new methods of data taking and data reduction, as described below. In addition, our UV and EUV observations utilize the constant seeing and 1-arcsecond resolution of *TRACE*. Not only chromospheric but also eclipse coronal imaging have recently used new image-processing methods to improve available spatial resolution (Pasachoff et al., 2008; Pasachoff, 2009).

## 2. Observations

### 2a. Swedish 1-m Solar Telescope

The *Swedish 1-m Solar Telescope* of the Institute for Solar Physics of the Royal Swedish Academy of Sciences is a vacuum 1-m telescope located on La Palma, Canary Islands, Spain, where it opened in 2002. Its adaptive optics requires a two-dimensional lock, essentially requiring a sunspot or pore in the field of view (or at least well-defined granulation), which is difficult to have near the solar limb.

Our four-day observing run ranged from 11 to 14 July 2006, and was coordinated with *TRACE* observations of the same solar region. The best seeing was on 13 July and 14 July. We selected about an hour of the best seeing from each day plus samples of high-quality seeing from other times, 11:43:00 to 12:43:00 UTC and 14:09:00 to 14:13:00 UTC on 13 July and 07:57:00 to 08:59:58 UTC and 14:10:27 to 14:12:51 UTC on 14 July.

The Hα observations were made using three synchronized Sarnoff CAM1M100 1k×1k CCDs. Ten percent of the light from the 1.7 nm FWHM pre-filter was diverted by a beam-splitter to two wideband cameras in a Phase Diversity configuration (one of them intentionally out of focus by about 12 mm). The remaining 90% of the light went through the Solar Optical Universal Polarimeter (SOUP) filter, owned by the Lockheed Martin Solar and Astrophysics Lab but on long-term loan to *SST*. It had first been used on Spacelab 2 flown aboard the space-shuttle *Challenger* in 1985. It is a temperature-compensated, servo-adjustable Lyot filter, with full-width at half maximum of 0.0078 nm in its narrow mode and 0.0128 nm in its wide mode, after a 0.8 nm prefilter to eliminate adjacent peaks. Wide-band Ca II H-line observations were also made, using Megaplus II cameras from Redlake.

We observed at Hα line center and at ±0.035 nm and ±0.070 nm. SOUP, under computer control, cycled through the tunings in sequences of 49.155 s. The tuning was such that the actual wavelengths were actually 0.0026 nm less than Hα during the July 13 observations and 0.0029 nm less on July 14 (Löfdahl, 2008). (The spectral lines are shifted by gas motion and solar rotation—the latter particularly at the limb, where we observed—so there was a calibration step where we scanned through the line close to the coordinates where we took data, resulting in our measurements of the offsets.) At each tuning, 122 frames were taken in 9.831 s, a rate of over 12 Hz, with exposure time of 17.2 ms, useful for keeping up with the 10-ms timescale of seeing changes.

Such tuning sequences, with images from all three cameras, were co-processed for correcting atmospheric phase aberrations by use of Multi-Object Multi-Frame Blind

Deconvolution (MOMFBD; van Noort, Rouppe van der Voort, and Löfdahl, 2005ab; see also Löfdahl, van Noort, and Denker, 2007). Camera misalignment was calibrated by use of images through a pinhole array, resulting in precise alignment of all the restored images in a MOMFBD set, in the wideband as well as in the different narrow-band SOUP tunings. Rouppe van der Voort et al. (2005) report obtaining data, with both adaptive optics (which we were unable to use) and MOMFBD, close to the 0.1 arcsec diffraction limit of the telescope.

After MOMFBD processing by Löfdahl with a computer cluster in Stockholm, we had one compound image every 10 seconds from each of the five wavelengths plus Dopplergrams for ±0.35 nm and ±0.070 nm from the center of Hα. Our processed images included 69 such sequences from July 13 and 72 from July 14. The subfield size used for MOMFBD processing was larger than usual, 256×256 pixels or 16.6×16.6 arcsec. (MOMFBD processing is applied separately to overlapping subfields, and mosaicking is seamless.) This subfield-size is much larger than the isoplanatic patch size but we found that restoration of smaller subfields often failed completely when they were completely off-disk, where the wideband instensity is very low. With the larger subfields, all subfields with off-disk features also include some part of the disk, where wideband phase diversity is more effective. The restored images were stored in floating-point format. Residual drifts were corrected later at Williams College using the position of a disk active region that appeared constant during our 14 July 2006 run (Figure 1).

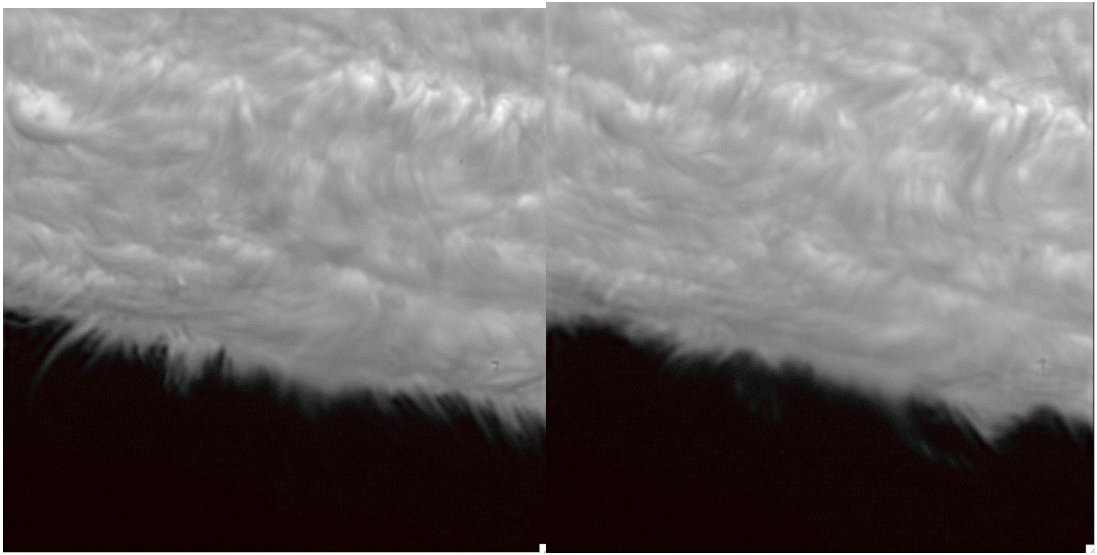

**Figure 1** *Two MOMFBD-restored core Hα images taken 14 July 2006 (07:57:00 UTC, left; 08:10:06 UTC, right). Though limb features changed substantially in the 13-m interval, we aligned on the disk emission at upper right.*

*2b. Transition Region and Coronal Explorer*

NASA's *Transition Region and Coronal Explorer* (*TRACE*) was launched into a sun-synchronous orbit on 2 April 1998. It has 0.5 arcsec pixels, giving a spatial resolution of 1 arcsec, and a field of view of 8.5×8.5 arcmin (Golub and Pasachoff, 2009; Handy et al., 1999). It has a 1024×1024 pixel array.

We observed mainly in the 160 nm continuum, which has a bandpass filter of 27 nm and includes C I and Fe II lines but mainly continuum, showing a temperature range of 4,000–10,000 K.

We also observed in the 17.1 nm emission line of an Fe IX/Fe X blend with a bandpass of 0.64 nm, resulting from a 160,000–2,000,000 K temperature range. Spicules show as dark silhouettes against the coronal continuum. Exposure times had to be at least 30 s and preferably at least twice that, with the longer exposure times (mostly substantially longer than the 160 nm exposures of 9.7, 19.5, or 27.6 s) desirable for image quality.

For some of our joint observations at 160 nm, *TRACE* observed with a reduced field of view of only 768×768 pixels to improve the cadence by reducing the readout time. A table of exposure times and observations is available in Jacobson (2008). During our prime 13 July 2006 run at SST, we have *TRACE* observations at 17.1 nm mostly with 32.8 s exposures and a cadence of 40 s, and a few with 46.3 s exposures with a 54 s cadence, and during our prime 14 July 2006 run at SST, we have a few *TRACE* observations at 17.1 nm with 55 s exposures and 63 s cadence followed for some (11 images) of the longer run with 160 nm observations with 9.7 s exposures at 15 s cadence and then most of the run (108 images) with 160 nm observations with 19.5 s exposures at 25 s cadence.

### 3. Analysis of Hα spicule images

*3a. Images and methods of analysis*

We compiled detailed statistics by measuring 40 spicules, 20 each from the 13 July 2006 and the 14 July 2006 SST runs. The spicules for which statistics were reported all clearly appeared to persist from frame to frame, after painstaking spatial alignment. Finding and following features sufficiently isolated or unambiguous over essentially their entire lives was extremely challenging, and so we were forced to restrict our study to the 40 presented events (Jacobson and Pasachoff, 2008). We have additional data in collaboration with SUMER on SOHO (Tingle and Pasachoff, 2008).

Next we discuss our methods for measuring spicule properties, the results of which are presented in Section 3b. The vertical velocity of each spicule examined from its apparent motion across the plane of the sky, assumed constant (with no better assumption available, given our cadence of about 50 sec, which itself was the highest possible to obtain with our setup), was found by dividing its height by the product of the number of frames in which it was ascending and the cadence. Each spicule typically was visible for 100 to 150 s, rising from the lowest points at which they could be seen to their maximum heights. In most cases, only ascending velocities were found because the majority of the 40 spicules measured seemed to disappear at their maximum heights, either just as they reached those altitudes above the limb or several frames later (Figure 2).

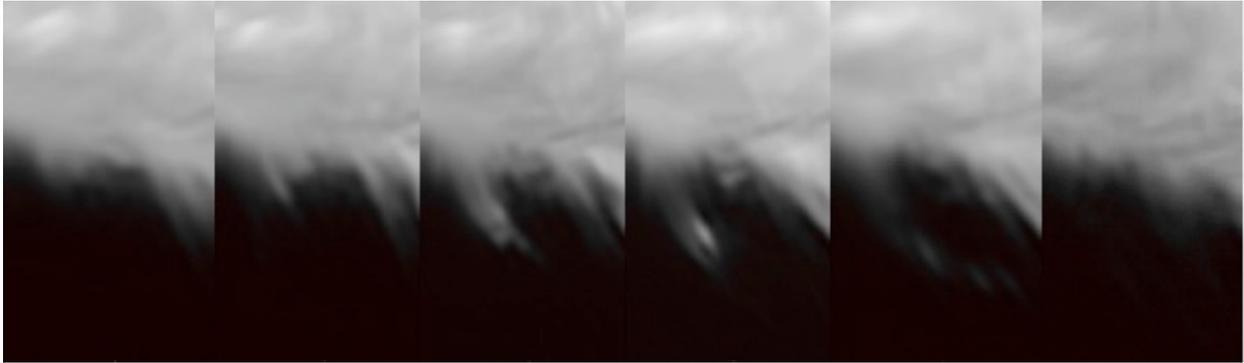

***Figure 2***  *The evolution of an H α spicule, seen at line center, visible in the left half of each frame as rising for four frames at constant intervals of about 50 s. The feature appears to disappear at its maximum height without visibly descending.*

Diameters were measured as the visible widths of jets at the midpoints of their heights when they were at their maximum extents. Line-of-sight velocities were measured by fitting Gaussians to 5-point spectra from the five wavelengths observed: ±0.070 nm, ±0.035 nm, and line center, all slightly offset as described earlier (Figure 3). Such line-of-sight velocities measured from the Doppler effect are more likely to measure true velocities than upward or downward velocities measured from motions of images along the plane of the sky, since the latter could reflect ionization fronts rather than actual motions. (Note that downward in our limb observations—along a solar radius at the limb—is a direction perpendicular to downward for on-disk observations of mottles.) Still, to measure the Doppler velocities from our 5-point spectra, we assume that the spicules are unsaturated, in at least two of our observed wavelengths, and that we are seeing the same feature rather than some combination of features including contributions from parts of objects in front of or behind the feature being examined. In any case, we see the same differences in every spicule measured, so we conclude that optical thickness does not invalidate these differences in velocity.

For our purposes at the limb, the base of the spicules is the location of the observed limb at Hα line center. Since we are observing in Hα, as opposed to Ca II H or K, the double-reversal of radiative-transfer does not dominate line center. However, since we are looking at a projection onto the plane of the sky, some spicules will be closer to us than the spicules we see at the sun's apparent rim, and some spicules will be farther, so there is no way to know what the actual base is of individual spicules we measure or therefore what the height on such a spicule is relative to its actual base as opposed to its apparent base.

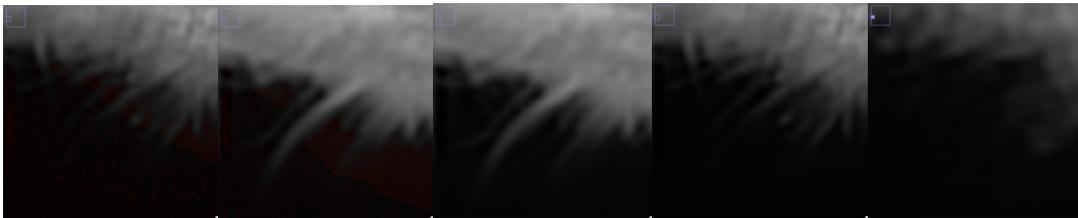

***Figure 3***  *Line-of-sight velocities were measured by fitting Gaussians to 5-point spectra using the intensity at maximum height in each of the images shown, which are separated by about 50 s*

*in time.   Here they correspond to shifts from Hα line center of –0.0729 nm, –0.0379 nm, –0.0029 nm, +0.0321 nm, and +0.0671 nm, going from the left to the right panel.*

Dopplergrams also show the velocities (Figure 4).

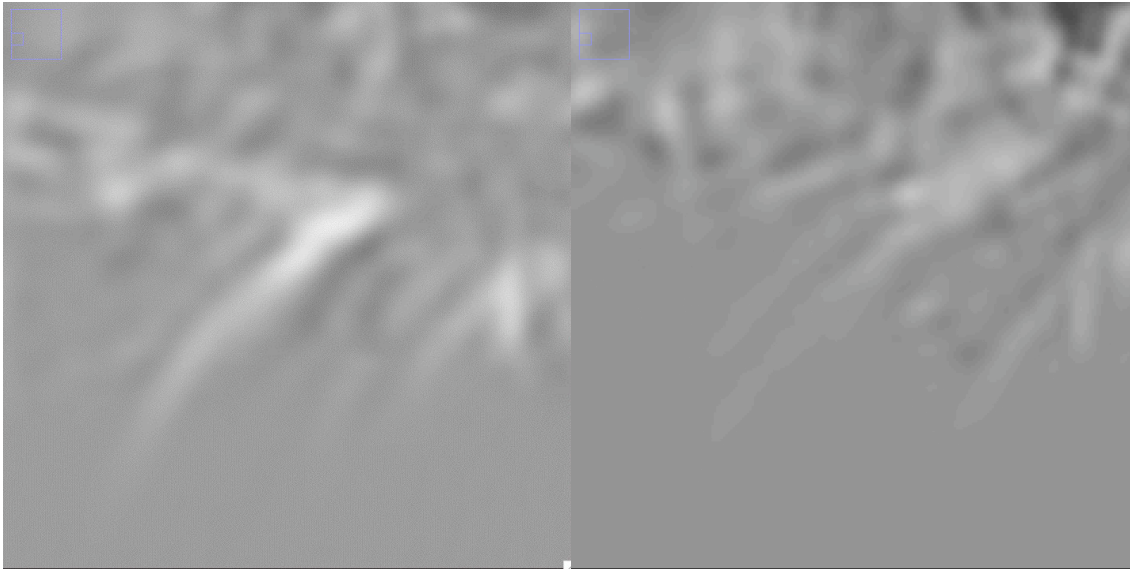

***Figure 4***  *Dopplergrams made with the –0.0379 nm and +0.0321 nm images from Figure 3 (left) and the –0.0729 nm and the +0.0671 nm (with respect to Hα line center) images (right).*

It is worthwhile to consider the quality of the EUV observations, since we are measuring small-scale features.   Though DeForest, Martens, and Wills-Davey (2009) have recently discussed diffraction in the TRACE telescope, those measurements do not apply at our continuum wavelength of 160 nm.   (The diffraction, rather than scattering, is caused by support mesh on the thin EUV filters that act as a grating, as described in Lin, Nightingale, and Tarbell, 2001; these filters are not used in the UV channel.   There is no measurable scattering in any of the TRACE channels, outside of special circumstances such as the transits of Mercury and Venus [for which see Schneider, Pasachoff, and Golub, 2004] that allowed the low level to be seen [Golub, 2009].)   We assume that contamination of the 160 nm continuum by C IV at 155 nm (at 100,000 K) and Fe II and He (formed at chromospheric temperatures) does not greatly affect our measurements.   Assessing the relative contributions of C IV, Fe II, and He to the 160 nm TRACE passband for limb features would have to await separate studies and measurements. Some on-disk analysis of assessing relative contributions by also observing the TRACE 170 nm band, which excludes the C IV lines at 154.8 and 155 nm, was carried out by de Wijn and De Pontieu (2006).

### 3b. Spicule statistics from the Swedish 1-m Solar Telescope data

The diameter measurements showed a distribution with a mean of 660±200 km with a median of 620 km, and a range of 300–1100 km (Figure 5). Both mean and median are close to the mean of 615 km of Nishikawa (1988), though, unlike his distribution, ours appears double-peaked, though the range of 3 to 10 spicules in each bin around the average size could result from small-number statistics. (The bin size was chosen in round numbers, but there is no indication that other bin sizes for the particular 40 spicules measured would have changed the shape of the distribution.)

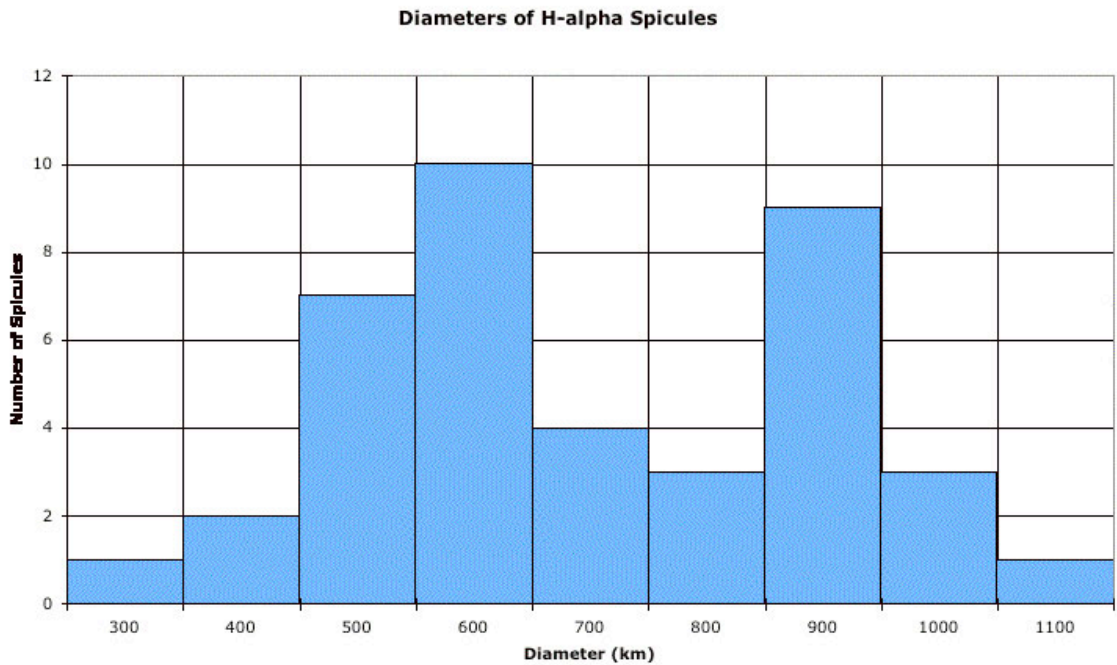

***Figure 5*** *A histogram showing distribution of diameters among 40 spicules.*

The maximum-height measurements showed a distribution with a mean of 7,200±2,000 km with a median of 6,900 km and a range of 4,200–12,200 km (Figure 6). The values match those found much earlier by Lippincott (1957) and others. Lippincott used 1000-km bin sizes for spicule heights; we verified that changing our 500-km bin sizes to 1000-km bins does not drastically change the shape of the distribution we measured. If we evaluate error bars in Poisson statistics, then the standard error of an individual bar is $\sqrt{[n*\bar{p}(1-\bar{p})]}$, where $\bar{p}$ is the number of counts on a category divided by the total number of counts $n$. So our highest bar, with 10 counts out of a total of 40, has a standard error of 2.7 and the bar with 4 counts has a standard error of 1.9.

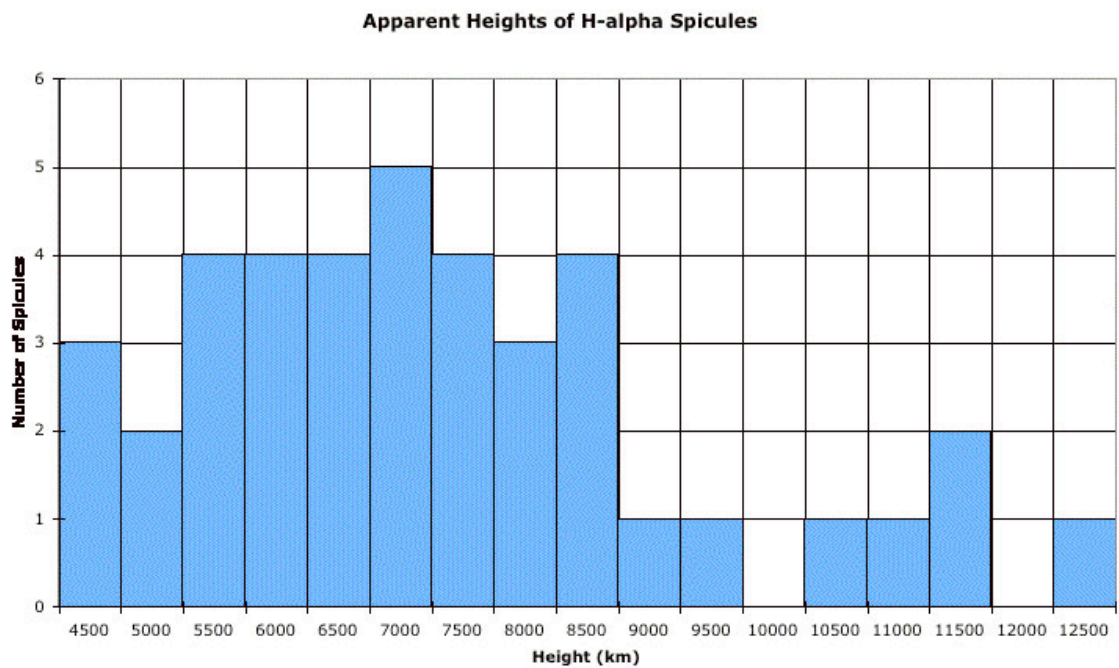

*Figure 6   A histogram showing distribution of maximum heights of Hα spicules. As a test, we also graphed with 1000-km bins; the overall shape of the bars' envelope did not change substantially.*

As measured by Heristchi and Mouradian (1992), the average spicule tends to lean toward the equator.  We were observing slightly north of the on the west limb, and we found the mean and median tilt of the spicules in a southern, equatorial direction from the normal, though within statistical uncertainty (Figure 7). In any case, our spicules were not measured to tilt as strongly as the 14° mean tilt reported by Heristchi and Mouradian. Our values appeared to have a tilt of 1.4°±27.7° with a median of 5.0° and a range of −55.0° to 50.0.  Correcting to absolute values (that is, the absolute value of the angle between the axis of a spicule and the line normal to the solar surface at the equator, to eliminate the insignificant point of whether spicules are inclined to the apparent left or right (i.e., the + or − direction) from our point of view, gives 23.4°±15.0° with a median of 23.0° and a range of 0.0° to 55.0.  (We also plotted the ± angles to determine if there was any correlation between the direction in which spicules were leaning and the direction toward the solar equator, which we did not find.)

 Heristchi and Mouradian's measurements were taken in 1970, near solar maximum, while our measurements were taken in 2006, near solar minimum, and the shape of the magnetic field, of course, varies over the solar-activity cycle.  It would be of interest to consistently measure the inclinations of spicules over a solar-activity cycle, and to compare the distribution with the expected magnetic field geometry in the overlying corona.

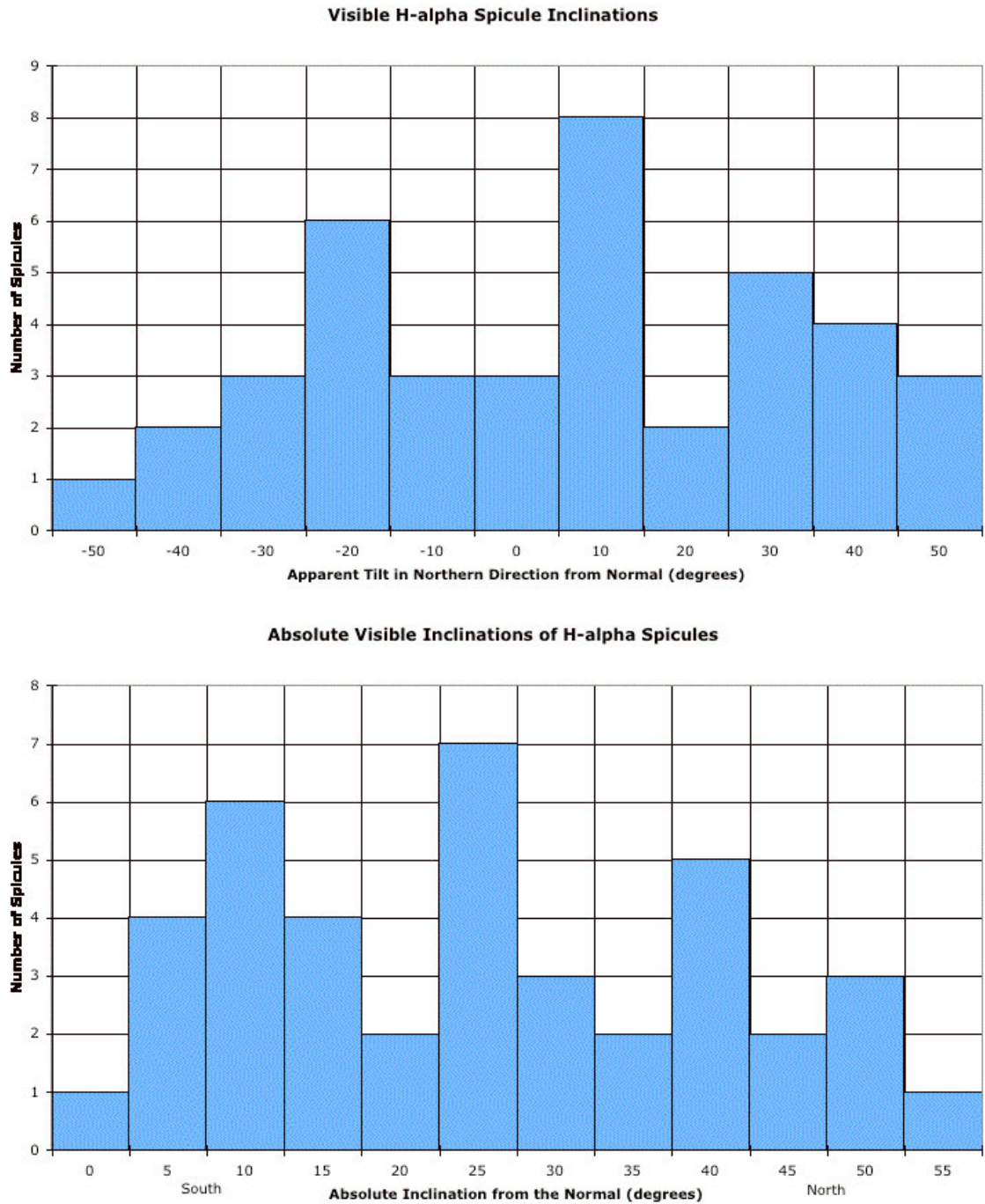

***Figure 7***  *Histograms showing (a) distributions of visible inclinations and (b) absolute values of those inclinations.  Negative values indicate a jet tilted south from the normal.*

For spicule lifetimes, we found a mean of 7.1±2.3 min with a median of 7.0 min and a range of 3 to 12 min (Figure 8).  The lifetimes were, on average, about two minutes longer than the means calculated by Lippincott (1957) and those summarized by Beckers (1972), perhaps because of improved visibility at lower heights because of the improved resolution of the SST observations. Our improved resolution might also have allowed us to see spicules at their maximum heights for longer times.

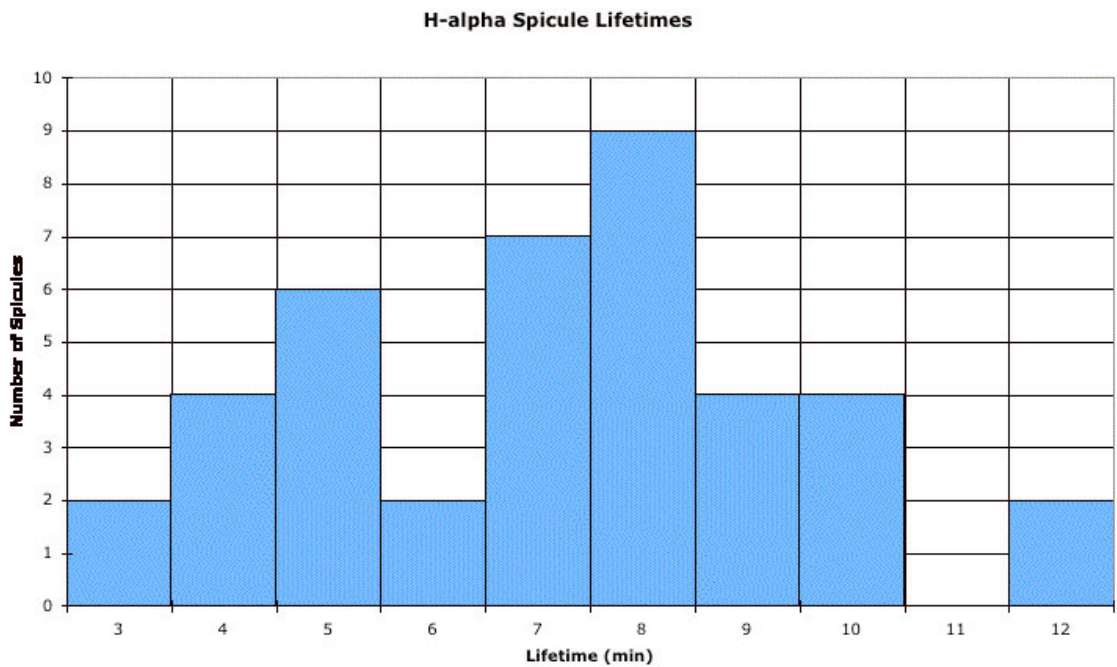

***Figure 8*** *A histogram showing the distribution of lifetimes, including not only the time it took them to travel from their minimum heights to their maximum heights (as listed in Section 3a) but also the time they seemed to be visible at their minimum heights prior to rising (an apparent stationary phase), the time at which they remained at their maximum heights, and the time it took for them to fall from their maximum to minimum heights. The apparent stationary phase was an impression gained from careful looking at the individual spicules, which seemed to be obviously visible for one or more frames before they were seen to rise.*

The spicule ascending velocities we measured have a mean of $27.0 \pm 18.1$ km s$^{-1}$, a median of 25.0 km s$^{-1}$, and a range of 3.0 to 75 km s$^{-1}$ (Figure 9). These values are consistent with the commonly accepted value of 25 km s$^{-1}$ (Beckers, 1968, 1972).

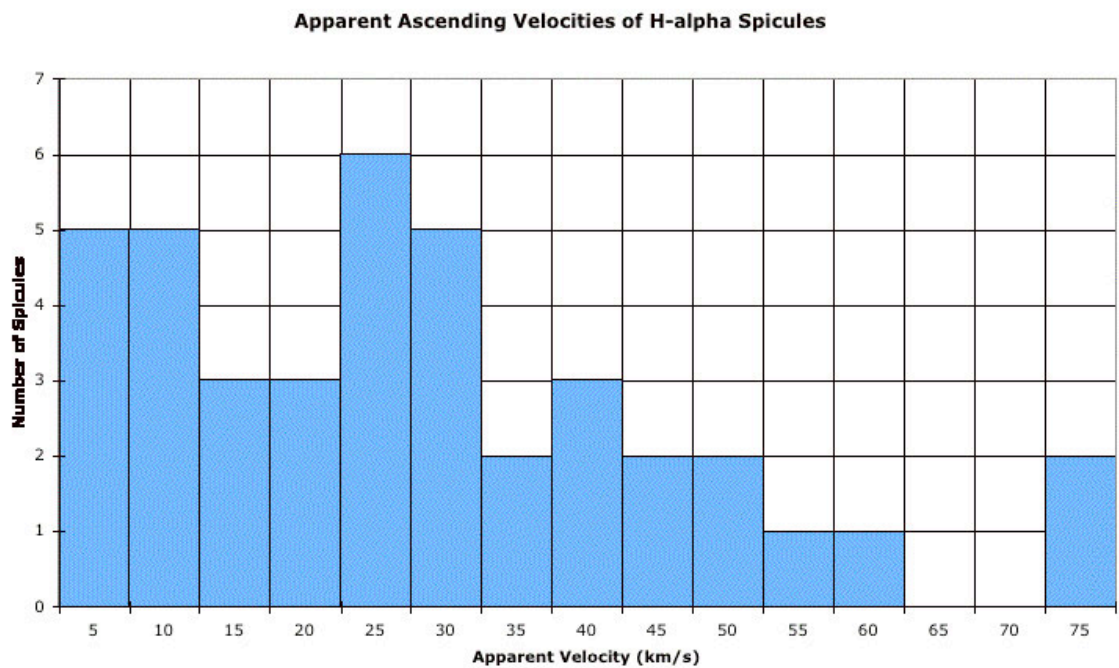

***Figure 9*** *A histogram showing the distribution of apparent ascending velocities.*

Only 11 of the 40 spicules for which we have statistics did not fade out at their maximum heights, so our statistics for descending velocities are insufficient, with a mean of 29.4±8.7 km s$^{-1}$, a median of 29.0 km s$^{-1}$, and a range of 14.0 to 42 km s$^{-1}$ (Figure 10). Nevertheless, our values are consistent with previous studies, which found similar velocities of ascent and descent.

***Figure 10*** *A histogram showing the distribution of apparent descending velocities among the 11 Hα spicules that did not disappear at their maximum heights.*

We measured line-of-sight velocities for the spicules when they were at their maximum heights from Doppler shifts calculated from 5-wavelength sets of images. We measured a directional mean of −2.6±6.7 km s$^{-1}$, a median of −2.2 km s$^{-1}$, and a range of −30.2 to 9.9 km s$^{-1}$. Correcting to absolute values gives 5.1±5.1 km s$^{-1}$, a median of 3.8 km s$^{-1}$, and a range of 0.3 to 30.2 km s$^{-1}$ (Figure 11). Our absolute mean agrees with the value of 6 km s$^{-1}$ quoted by Beckers (1972). If our single outlier with a line-of-sight velocity of −30.2 km s$^{-1}$ is omitted, then our mean becomes 4.3±3.2 km s$^{-1}$.

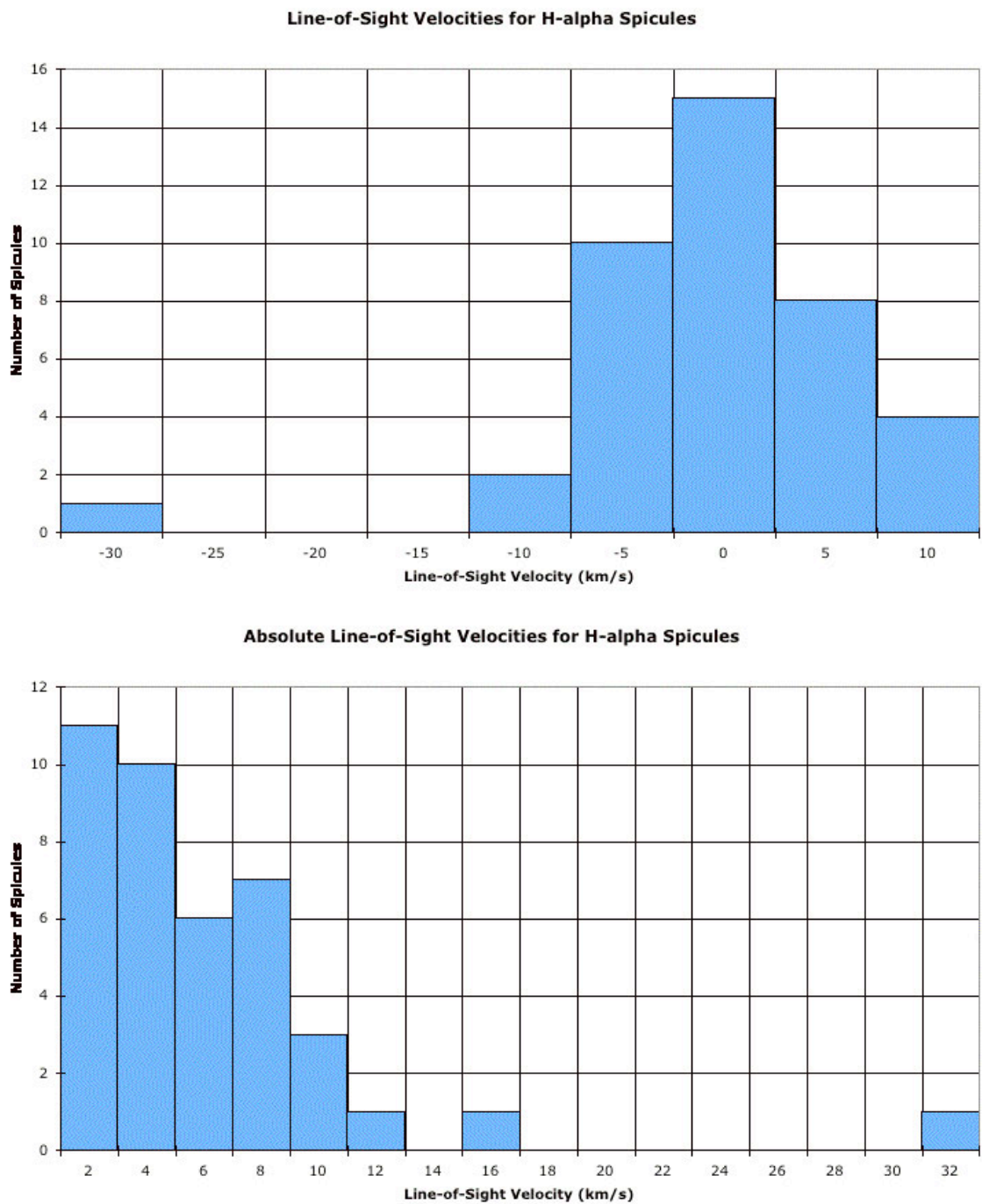

*Figure 11*  *Histograms showing distributions of line-of-sight velocities and absolute values of the line-of-sight velocities at maximum jet height, that is, the tops of the spicules when they were at their maximum heights.*

For the distributions of line-of-sight velocities at the bases of the 40 spicules measured (we defined the base for each day's observations as the location of the apparent limb at line center, and used the same location for all spicules measured from that day) we find $3.1 \pm 11.9$ km s$^{-1}$ with a median of 1.3 km s$^{-1}$ and a range of $-17.8$ to 51.2 km s$^{-1}$ (Figure 12). For the absolute differences in line-of-sight velocities between the maximum heights and bases of these spicules, we find a mean of $11.9 \pm 11.3$ km s$^{-1}$, a median of 1.5 km s$^{-1}$, and a range of 0.4 to 55.2 km s$^{-1}$.

Pasachoff, Noyes, and Beckers (1968) and Nikol'skii and Sazanov (1967) found that line-of-sight velocities are not highly dependent on the height of plasma within a spicule within about 5,000 and 6,800 km and 4,500 and 7,500 km, respectively, above the solar surface. Tsiropoula, Alissandrakis, & Schmieder (1994), and Tziotziou, Tsiropoula, and Mein (2003, 2004) found that mottles they observed exhibited bi-directional flows, with the base portions showing downward motions while the tops of the structures had alternating upward and downward motions. They argued that these motions could be explained by magnetic reconnection, such as in the picture due to Pikel'ner (1969). From our Doppler measurements, we regularly found oppositely directed motions at lower levels in the 40 spicules we examined, consistent with the results of the above mottle papers. The plasma motions at footpoints were measured to have approximately the same absolute mean—$3.1 \pm 11.9$ km s$^{-1}$—as that of the often oppositely directed motions near the maximum heights of each jet, which averaged $2.6 \pm 6.7$ km s$^{-1}$. As shown in Figure 12, the absolute difference in velocities between the two plasma flows had a mean of $11.9 \pm 11.3$ km s$^{-1}$, which is significant considering that it is greater than the sum of the average magnitudes of the two components.

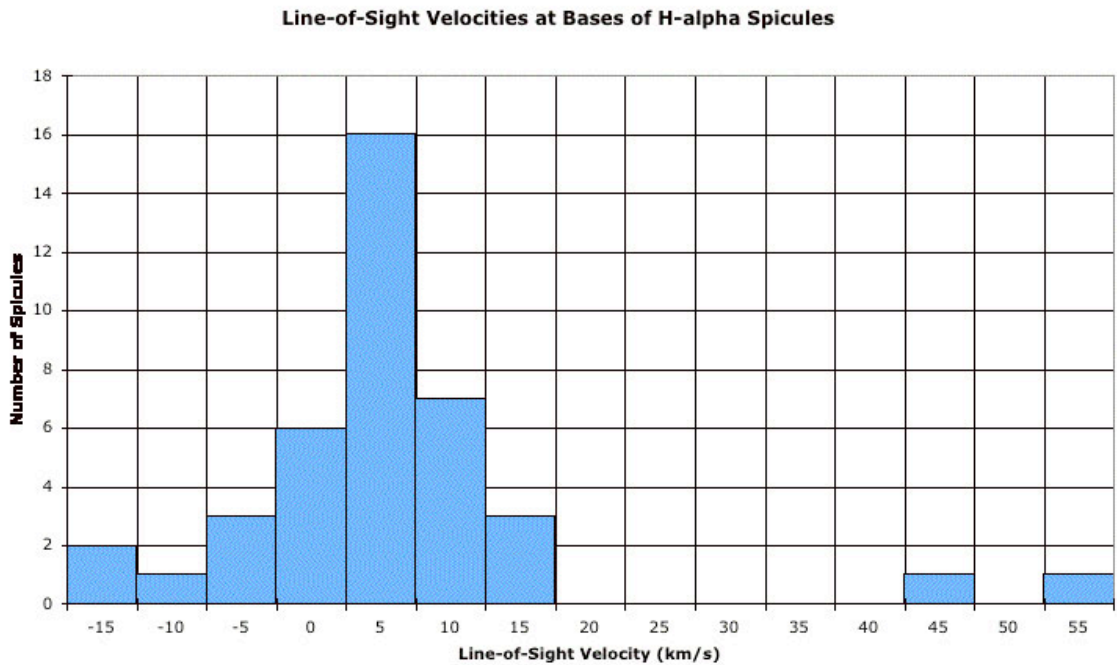

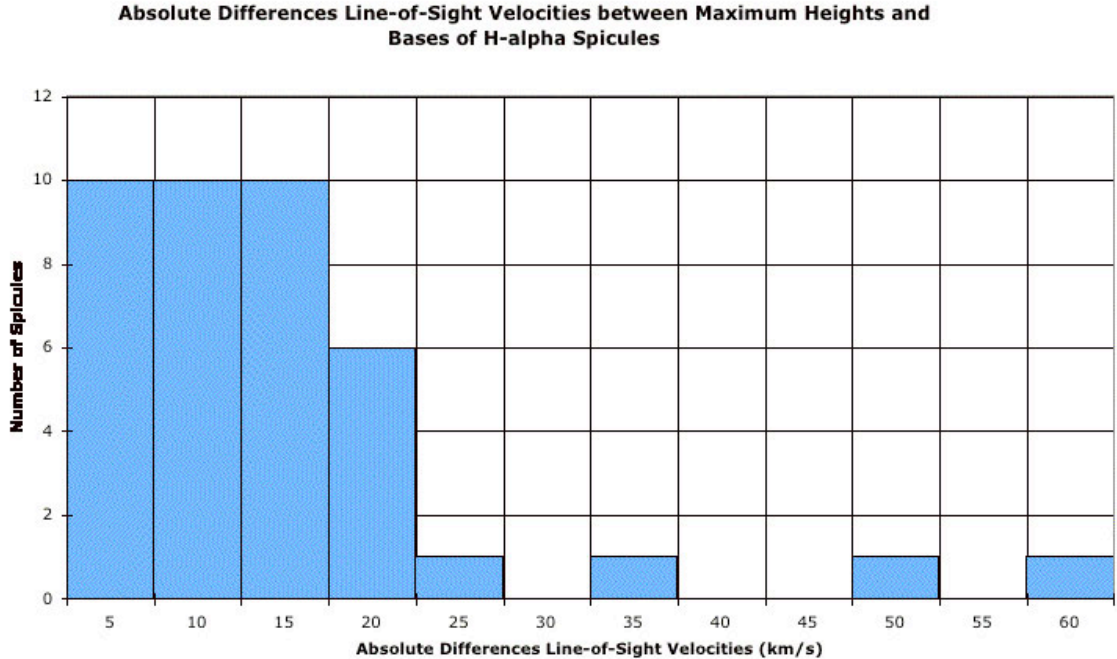

**Figure 12** *(top) The distributions of line-of-sight velocities at the bases of the 40 spicules observed in Hα. (bottom) The distributions of absolute values of differences in line-of-sight velocities between the maximum heights and the bases of these spicules.*

### 3c. Spicule statistics from the TRACE data

We coaligned the *TRACE* data with the *Swedish 1-m Solar Telescope* data, a nontrivial task given the differing image scales (0.0518 arcsec/pixel for SST and 0.50 arcsec/pixel for *TRACE*), and the different cadences. Though we could identify the Hα spicules that we saw with the *SST* with the spicules at 160 nm (that is, chromospheric emission) that we observed with *TRACE*, we were unable to identify definitively spicules observed with *SST* in Hα with the spicule-shaped silhouettes that we observed at 17.1 nm (that is, coronal emission) with *TRACE,* which would have mimicked the correlation of *TRACE* and *SVST* (Swedish Vacuum Solar Telescope, a precursor in the same tower of *SST*) images by Berger et al. (1999a). He also saw the spicules as silhouettes—absorption of chromospheric gas against a coronal background at that wavelength. Unfortunately, our *TRACE* 17.1 nm cadence was so much slower than our SST Hα cadence that, coupled with the different image scales, we could not make satisfactory cross-identifications.

As we had for the *SST* data, we defined the bases of spicules in the *TRACE* data as the points of maximum intensity along axial lines drawn through the center of each selected spicule. After MOMFBD processing, the *SST* images show their superior resolution to the *TRACE* images (Figure 13).

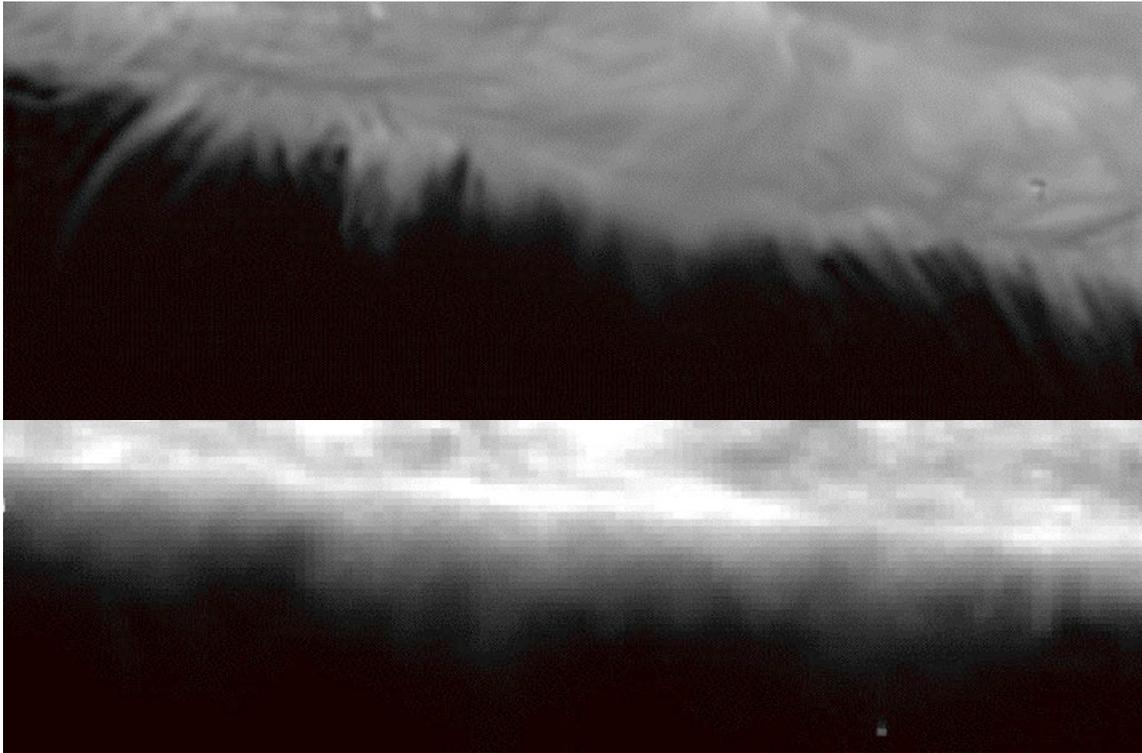

***Figure 13*** *Coaligned images of the west limb taken within about one second of each other by SST in Hα (top) and TRACE at 160 nm (bottom) on 14 July 2006. Even with the substantial difference in resolution, some features are identifiable on both.*

We identified 14 common spicule-like structures in both *SST* Hα and *TRACE* 160 nm datasets, using the computer program ImageJ. We found (160 nm)/Hα statistics for spicule diameters to be 1610±430/670±180 with medians of 1,600 km/620 km and ranges of 700–2,500 km/360–900 km (Figure 14). Since the TRACE resolution (2 pixels) is 1 arcsec = 726 km, essentially all the measured widths in Figure 14 are resolved.

(Since we could unambiguously align individual features that resided well above the limb, the different heights of formation of the continuum at Hα with SST and 160 nm with TRACE should not have caused a problem, other than a selection effect.) As shown in Figure 14, there is essentially no correlation ($R^2$=0.006) between the two values of diameter for SST and TRACE spicules. However, in every case the width measured with *TRACE* at 160 nm was greater than the width measured with *SST* at Hα by a factor of over 1.5, with this value exceeding 2.5 for the medians,—. The *TRACE* widths we measure are impacted by the telescope's broad point-spread function (DeForest et al., 2009). Moreover, the resolution of *SST* with MOMFBD is approximately four times finer than that of *TRACE*, though both should have been adequate to make the measurements reported. Nonetheless, our observations indicate that relatively hot plasma emitting in the UV extends further from the axis than the cooler material that corresponds to Hα.  This implication is consistent with Dere, Bartoe, and Brueckner (1983), who determined that chromospheric jets observed in wavelengths between 117 nm and 171 nm have diameters between 1 and 4 times larger than that of correlated Hα spicules.

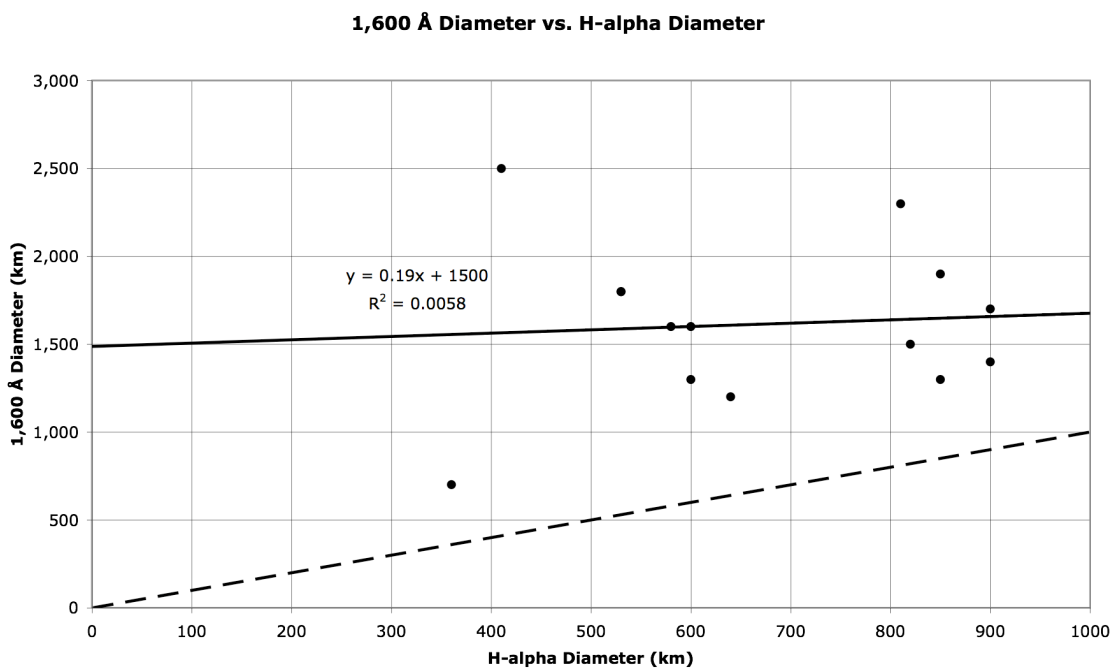

**Figure 14** *A graph of diameters in 160 nm from TRACE vs. H α diameters from SST for the 14 spicules that were detectable in both data sets. The dashed line shows what a 1:1 correspondence between the diameters at the two wavelengths would be.*

As was true for spicule diameters, the maximum heights measured were in all cases greater at 160 nm than in Hα (Figure 15).  Values for 160 nm (Hα) were 10,300±2,700 km (7,400±2,300 km) with medians of 10,600 km (7,100 km) and ranges 5,600 to 14,700 km (4,100 to 12, 200 km).  The heights found at the two wavelengths were strongly correlated ($R^2$=0.85). The regression line shows that, on average, the altitude reached in the solar atmosphere in the UV continuum wavelength is about 2,800 km greater than that reached in Hα.  Earlier, as yet unpublished, *TRACE* vs. *SST* (pre-MOMFBD) observations analyzed by Westbrook (2006) similarly found mean heights of about 8,000 km in Hα and 12,000 km in 160 nm for a different set of spicules.

Pasachoff (1969) discussed the volume of the 3-dimensional chromosphere-corona transition zones, which are characterized by observing ions like O V but which are often treated as plane-parallel.  Westbrook (2006) used the data collected by De Pontieu et al. (1998) to argue that the low number of spicules observed in O V – in spite of abundant overall emission in this wavelength – was an indication that the plasma in question did not compose the jets themselves. His resulting conclusion was also based on the statistics of EUV spicules determined in other studies, many of which showed that these structures generally have similar properties but slightly larger dimensions than their H α  counterparts.  For example, Xia et al. (2005) determined that, in the N IV line at 765 Å, spicule heights are at least 7,500 km, a value that falls in the commonly-accepted range for the average height of Hα spicules.  Generally, none of these EUV features reached altitudes more than 15,000 km above the limb, though Xia et al. (2005) did observe "long macro-spicules" as tall as 30,000 km.  Westbrook (2006) also cited Dere, Bartoe, and Brueckner (1983), who calculated that spicule diameter can be as much as four times greater in the EUV than in Hα.  Based on these statistics and the observations of De Pontieu et al. (1998), he concluded that sheaths of transition-region plasma may surround Hα spicules (Westbrook, 2006).  However, the importance of these sheaths, or if they actually exist, remains unclear.  Pasachoff (1969) had earlier posited that such sheaths would be necessary intermediaries between the chromospheric-temperature spicular gas and the coronal-temperature gas between at least the upper parts of the spicules, and calculated that the volume in such a three-dimensional transition zone must be substantial.

Such a sheath model predicts that chromospheric spicules would appear taller when viewed in the UV and EUV than in the optical.  The result is endorsed by the 2,900 km and 3,500 km differences in the means and medians, respectively, in our current data at 160 nm (untroubled by the diffraction seen in *TRACE* data at shorter wavelengths; there is no diffraction in the UV with *TRACE*), and by the similar difference of approximately 4,000 km found in Westbrook's study.

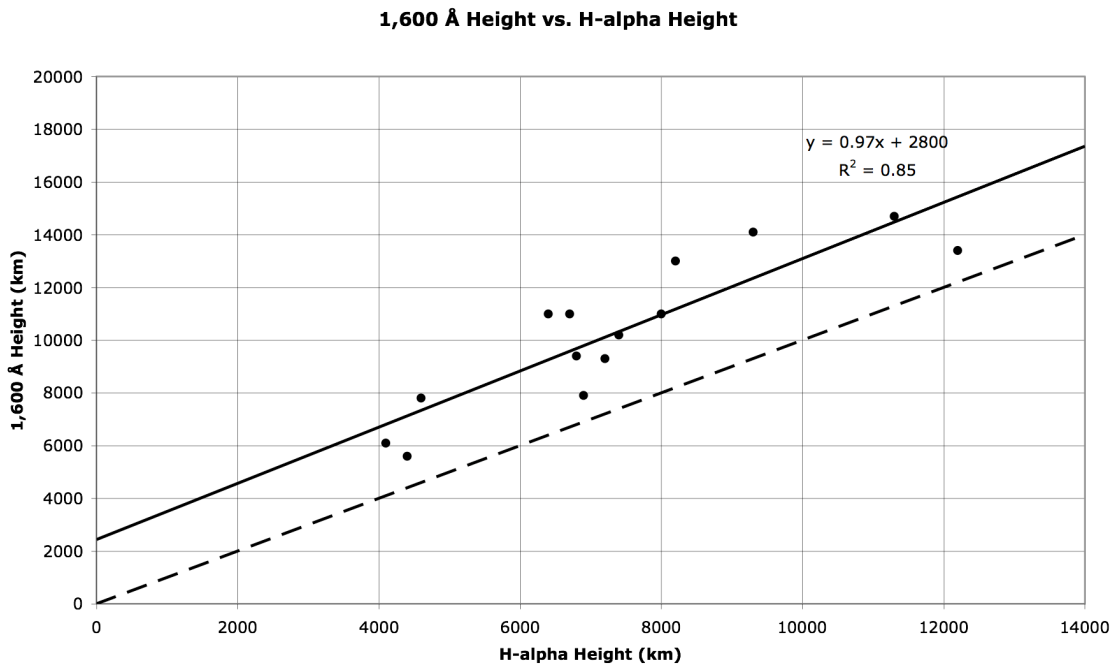

**Figure 15**  *A graph of heights in 160 nm from TRACE vs. Hα heights from SST for the 14 spicules that were detectable in both data sets.  The dashed line shows a 1:1 correspondence between the heights at the two wavelengths and the solid line is a fit.*

Our measurements of inclinations show 13.1°±29.2° for 160 nm (that is, commensurate with no inclination) with a median of 21° and a range of −30° to 51°, and 13.7°±29.0° for Hα with a median of 26° and a range of −32° to 50° (Figure 16). The inclinations observed in the two wavelengths are strongly correlated ($R^2$=0.94).

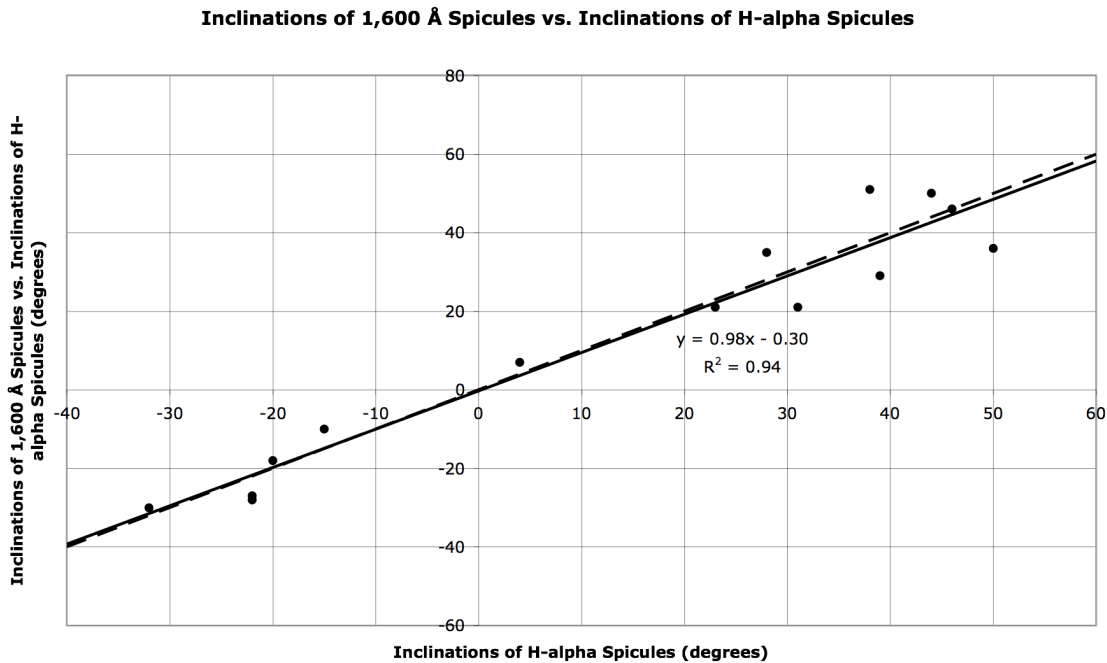

**Figure 16** *A graph of inclinations in 160 nm vs. inclinations in Hα for the 14 spicules in both data sets. A positive angle indicates a northward tilt from the normal line to the limb. The dashed line shows a 1:1 correspondence between the inclinations at the two wavelengths. The identical inclinations of individual features suggest that the features seen at 160 nm are affected by the orientation of the magnetic field as their Hα counterparts.*

The lifetimes measured at 160 nm were shorter than those found in Hα in 11 of 14 cases, 4.3±1.5 min at 160 nm vs. 7.3±2.4 min at Hα (Figure 17). The medians were 4.0 min/8.0 min with ranges 2 to 8 min/3 to 10 min for 160 nm/Hα, respectively. We cannot, however, rule out that some of the differences between the lifetimes in the two wavelength regimes could be due to differences in spatial resolution.

Though the relationship between the lifetimes of 160 nm vs. Hα spicules is consistent with the early EUV study of Dere et al. (1983), our mean and median values were not nearly as short as the 40 s they reported. This discrepancy may be the result of the relatively high levels of the solar atmosphere that corresponded to the wavelengths being observed in the Dere et al. study. Our group's early measurements (Kozarev, 2005) gave a mean half-lifetime of 3 min for both the rising and falling phases of an EUV spicule's lifetime. Since in our latest study, several 160 nm spicules were observed to descend after reaching their maximum heights, the Kozarev observations seem to be in reasonable agreement with our newer data.

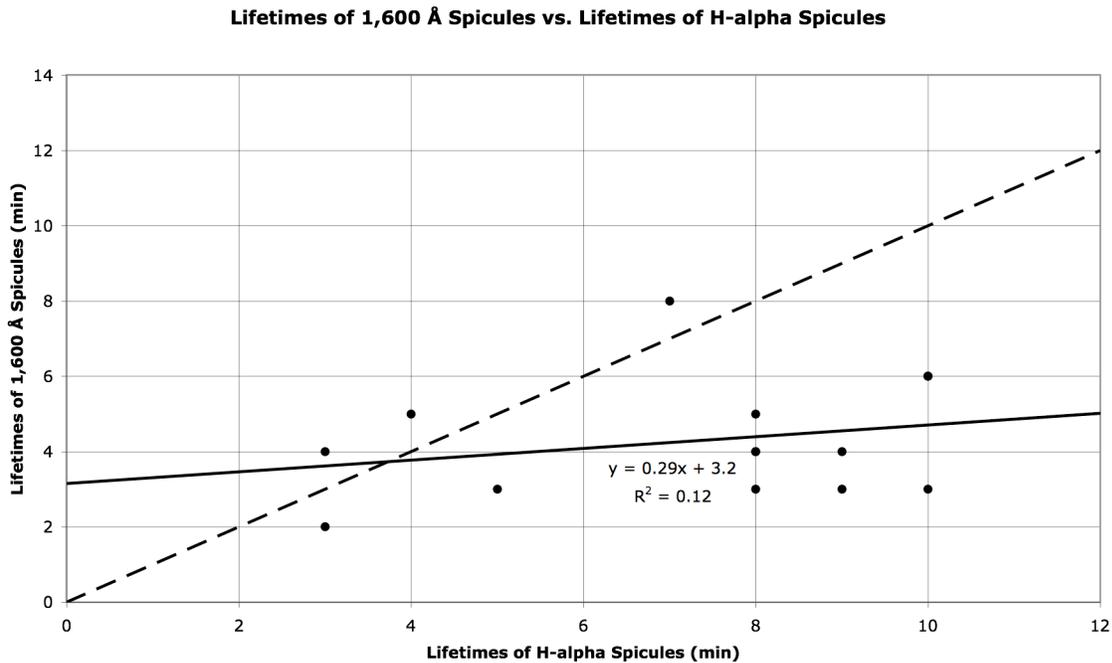

***Figure 17*** *A graph of lifetimes in 160 nm vs. lifetimes in Hα for the 14 spicules in both data sets. The dashed line shows a 1:1 correspondence between the lifetimes at the two wavelengths.*

The apparent ascending velocities measured were comparable for 160 nm/Hα, with values $34.6\pm15.8$ km s$^{-1}$/$31.4\pm16.0$ km s$^{-1}$, with medians 34.5 km s$^{-1}$/33.0 km s$^{-1}$ and ranges 12 to 65 km s$^{-1}$/3 to 58 km s$^{-1}$ (Figure 18). The values are strongly correlated ($R^2=0.75$) and the means and medians are essentially the same. Our earlier studies using *TRACE* data, Kozarev (2005) and Westbrook (2006), however, found mean velocity values of $12.9\pm1.1$ km s$^{-1}$ and $14.0\pm1.7$ km s$^{-1}$, respectively; these values are a factor of two lower for unexplained reasons. In any case, the near-equality of velocities in both wavelength in all three studies agrees with the observations of De Pontieu et al. (1998), who reported observing similar velocity structures in a single large jet observed in both Hα and O V at 62.9 nm in the EUV.

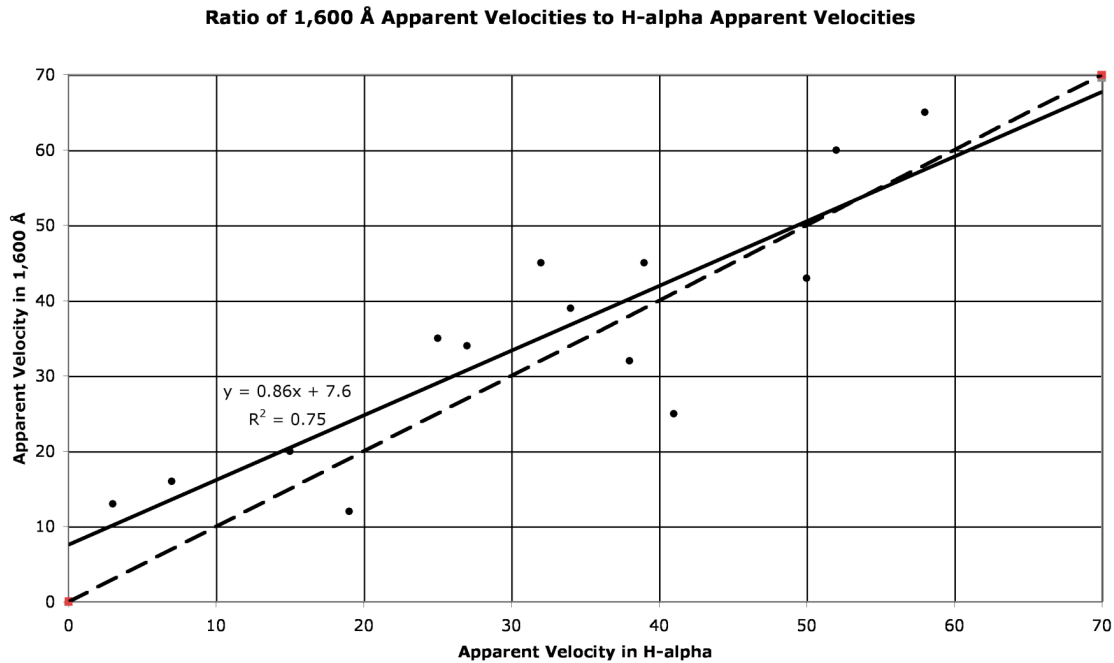

***Figure 18*** *A graph of apparent ascending velocities in 160 nm vs. apparent ascending velocities in Hα for the 14 spicules in both data sets. The dashed line shows a 1:1 correspondence between the apparent velocities at the two wavelengths.*

## 4. Conclusions

We were able to observe spicules at the limb with Multi-Object Multi-Frame Blind Deconvolution (MOMFBD) post-processing of images taken with the *Swedish 1-m Solar Telescope* and with the *TRACE* spacecraft (Figure 19). Most of the morphological and dynamical spicular properties that we determined for a sample of Hα features fit neatly into the framework of previous studies. Though higher than we found in preliminary work on this project, our means and medians for apparent velocity in both the 160 nm continuum and Hα agree with the generally accepted typical value of 25 km s$^{-1}$ for this statistic in Hα. Only a subset of the spicules we studied in both Hα (from the ground) and UV (from space) overlapped. Those that did had a slightly higher Hα velocity than the average (~35 km/s), and they had very comparable velocities in UV. This may imply that only faster spicules overlap in both wavelengths, but we cannot rule out a selection effect, given our relatively small number of features measured (though nonetheless a number that reflected lengthy work).

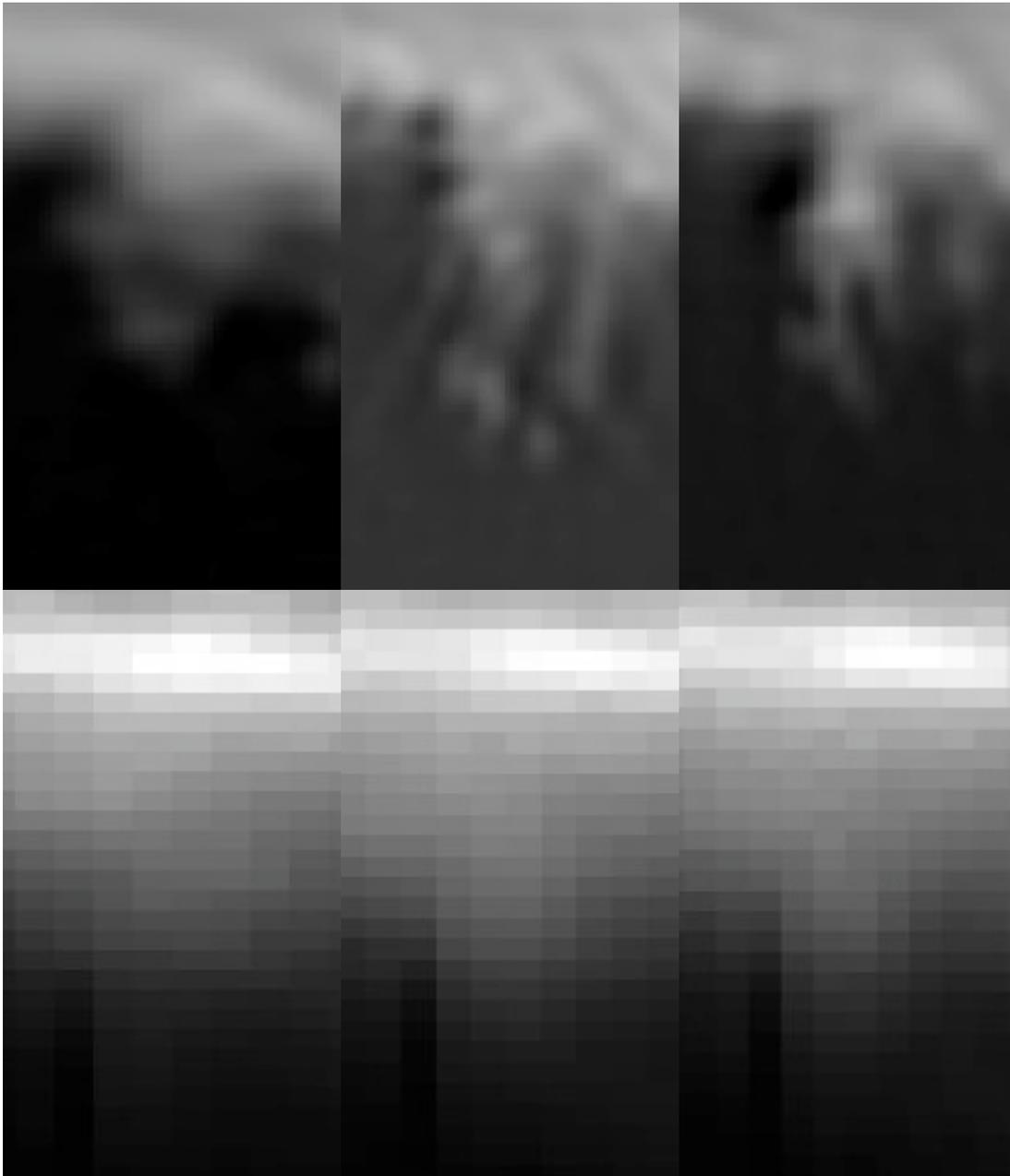

***Figure 19*** *(top) A spicule observed in Hα with SST on 14 July 2006, as it evolves over a 100 s period, with images at 50 and 100 s, respectively, after the initial image. (bottom) Images of the same field at the same times with TRACE in the 160 nm continuum. The MOMFBD post-processing of the SST images clearly improved its resolution over that of TRACE. Our work showed that the plasma emitting in the two distinct wavelengths moved at similar speeds.*

A detailed comparison of our findings with spicule models is beyond the scope of this paper. We will however discuss briefly implications of our results for some recent models. The structures and evolutions of spicules that we observed in Hα with *SST* were consistent with the *p-* mode oscillation leakage model of De Pontieu, Erdélyi, and James (2004) (the type-I spicules of De Pontieu et al., 2007a) in some ways, and with the magnetic reconnection model of mottles (Tziotziou, Tsiropoula, and Mein, 2003, 2004). Specifically, the 19 jets with lifetimes between 3 and 7 minutes and the 22 jets with apparent ascending velocities between 10 and 40 km s$^{-1}$—13 of which overlapped—behaved in the manner consistent with the properties given by De Pontieu et al. (2004). However, only 1 of the 13 spicules in this overlapping group had a height less than 1.l5 times the 4,000 km predicted by that model as formulated in De Pontieu et al. (2004); that is, most of our observed spicules are substantially taller than the heights predicted by the model. An important caveat with these comparisons however is that the $p$-mode model of De Pontieu et al. (2004) was matched with spicules seen in active regions, and its application to spicules in quieter regions, such as we observe here, is not known, but may result in taller spicules according to De Pontieu et al. (2004). Among the spicules that we observed at the limb to descend in Hα, their falling velocities (14–42 km s$^{-1}$) were consistent with the predictions of De Pontieu et al. (2004) for features on the disk; however, most limb spicules faded from Hα before falling, and the $p$-mode oscillation model does not account for such fading. Perhaps additional heating may be acting to cause our spicules to fade in Hα (e.g. Sterling & Hollweg 1984), although type I spicules are reported to not fade (De Pontieu et al. 2007a). There is no doubt, however, that we preferentially selected taller, easier-to-identify spicules in this study, and it has yet to be considered whether the $p$-mode oscillation model can explain such spicules occurring in appropriate (and in as-yet-undetermined) quantities. We also cannot rule out the possibility of that model applying to some, perhaps shorter, spicules, and a different model applying to other, including taller, spicules. Because of our rather coarse cadence, we are not able to determine whether the height-time trajectories of our taller spicules are parabolic, like type 1 spicules (De Pontieu et al. 2007a), or more complicated than parabolic (Hollweg 1982).

The fact that all our spicules appeared to have bi-directional plasma flows when at their maximum heights is consistent with the magnetic-reconnection scenario discussed by Tsiropoula, Alissandrakis, and Schmieder (1994) and Tziotziou, Tsiropoula, and Mein (2003, 2004), who saw similar behavior in mottles, though it might be consistent with other models as well. In addition, 29 of our 40 jets faded out at their maximum heights. Interestingly, the Type II spicules are also observed to fade, and magnetic reconnection has also been suggested as a possible mechanism for generating those features (De Pontieu et al. 2007a). Type II spicules, however, appear to be very different in their properties from "traditional" Hα spicules, and their lifetimes are much shorter, ~10-60 sec, and so we cannot hope to observe them as individual entities in the present study. Their relation with the "traditional" spicules remains a mystery.

In summary, our observed limb spicules: (i) seem to most closely resemble the motions of Tsiropoula, Alissandrakis, and Schmieder (1994) and Tziotziou, Tsiropoula, and Mein (2003, 2004) mottles, with the bi-polar flows consistent with a reconnection mechanism; (ii) do not closely resemble the properties suggested of p-mode oscillation (type-I) spicules (as seen in active regions on the disk); and (iii) cannot be directly compared with Type II spicules. *Hinode* also reveals other types of chromospheric jets for which the relationship with ground-based-observed traditional spicules is not yet clear (Shibata et al. 2007, Suematsu et al. 2007). Our data set, however, consisted of forty Hα spicules that were well observed, and 14 corresponding features seen in UV. Further differentiation between the models would come with increased sample size of our type of investigation. We look forward to further studies on all of these points.

Assuming that selection bias or statistical variations do not drastically shape our measurements, our SST statistics and observations, when combined with simultaneous *TRACE* data in the 160 nm continuum, appear to support a quiet-region model driven by reconnection. The larger heights and diameters in the UV wavelength are consistent with an overlying plasma layer surrounding and moving with an Hα jet, as are the equal inclinations and velocities of the gases emitting in the two lines. If it is assumed that this layer is a transition zone around the enclosed spicules, acting as the source of the 160 nm radiation, then the discrepancy in lifetime between jets viewed with SST and those observed with TRACE could be accounted for by the early period of energy transfer that occurs prior to any substantial UV emissions. The results of our project therefore are consistent with the sheath model; the volume in transition-zone sheaths is substantial (Pasachoff, 1969). See also Berger et al. (1999b). It would be useful to obtain 170 nm observations to rule out contributions of C IV to the EUV continuum imaging at 160 nm that we obtained.

## 5. Suggestions for the Future

The best way to improve spicular statistics is to obtain high-time-resolution observations at resolution of approximately 0.1 arcsec with the steady seeing from space. However, no such spacecraft imaging is planned. *TRACE* pixels, pre-CCD, are 0.5 arcsec, giving a resolution of 1 arcsec over 1/6 of the sun, and *Hinode X-Ray Telescope (XRT)* uses a 2000×2000 pixel CCD with 1 arcsec pixels though with full-sun coverage, and with higher resolution of 0.2 arcsec by using the *Solar Optical Telescope (SOT)* at the Ca II H-line (Suematsu and SOT Team, 2007; De Pontieu et al., 2007a; De Pontieu et al., 2007b; Tsuneta, et al., 2008). (Spicule images with *SOT* also appear in Okamoto et al., 2007; see also Pasachoff, 2007.) The *Hinode/SOT* observations, though, have 0.2 nm FWHM spectral resolution, compared with the 0.016 nm resolution we obtained at *SST*, ten to twenty times finer, so no Doppler information is available with *SOT* at the Ca II wavelength. The *Atmospheric Imaging Assembly (AIA)* instrument on *Solar Dynamics Observatory (SDO)*, scheduled for launch in late 2009, uses four 4000 ×4000 pixel CCD's, one on each telescope, to give 0.6 arcsec pixels and therefore 1.2 arcsec resolution over the whole sun.

With the *Swedish 1-m Solar Telescope*, we have obtained 0.2 arcsec resolution through the MOMFBD image processing of images taken at times of excellent seeing and with extremely high spectral resolution, 0.078 Å = 0.008 nm, using the Lockheed Martin Solar Optical Universal Polarimeter, SOUP (Berger, 2004). Still, at the limb we were unable to obtain a lock with the adaptive optics because of the absence of a pore or other two-dimensional object needed for the *SST*'s AO system. One first step would be a one-dimensional AO system, able to improve the imaging by locking on the limb. Future observations, at a time of more active sun, could potentially image spicules at a time when a pore or other silhouetted dark solar feature is available within the field of view near the limb; extended observational runs would no doubt be needed to find such occasions. Such 0.1 arcsec observations are potentially available at a handful of the best solar telescopes. The Big Bear Solar Observatory has installed (as of summer 2009) a 1.8-m solar telescope at its location that is noted for excellent seeing, but the telescope is not yet fully functional and its adaptive-optics system will follow later. The *Advanced Technology Solar Telescope (ATST)*, with its 4.24-m-diameter mirror, is not yet approved for construction; it is planned to have an integrated high-order adaptive-optics system that should give a resolution of 0.022 arcsec at 430 nm (Rimmele, Keil, and Wagner, 2006; Rimmele, Keil, and Dooling, 2008). But ATST would not be available until at least 2015, in time for the 50th anniversary of the Pasachoff, Noyes, and Beckers (1968) Sacramento Peak Observatory spicular observations of 1965.

## Acknowledgements

We thank Mats Löfdahl of the Royal Swedish Academy of Sciences for his work on Multi-Object Multi-Frame Blind Deconvolution of the *Swedish 1-m Solar Telescope* data and Michiel van Noort there for helpful assistance. We thank Evan Tingle (Keck Northeast Astronomy Consortium Summer Fellow from Wesleyan University, sponsored by a Research Experiences for Undergraduates grant from the National Science Foundation) for his collaboration with data reduction. We are grateful to C. Alex Young (NASA's Goddard Space Flight Center) and Daniel B. Seaton (formerly Williams College, then University of New Hampshire, and now Royal Observatory of Belgium) for consultation on alignment of images. We thank former Williams College students Anne Jaskot and Megan Bruck for their participation while obtaining the data at the *Swedish 1-m Solar Telescope* and Jennifer Yee (KNAC Summer Fellow from Swarthmore College) for work on earlier *SST* spicule data. Bruck also worked on arranging MOMFBD. We appreciate the earlier undergraduate thesis work at Williams College on *SST* and *TRACE* data of Kamen Kozarev (now at Boston University) and Owen Westbrook (now at M.I.T.**).** We thank Bart De Pontieu (Lockheed Martin Solar and Astrophysics Laboratory) for assistance during the first of the three SST data runs, and Rolf Kever and the Telescope Operators for their help also on site. We thank Leon Golub, Edward DeLuca, and Jonathan Cirtain (Harvard-Smithsonian Center for Astrophysics) for consultations on the *TRACE* data reduction. We thank Steven P. Souza of Williams College's Astronomy Department for computing assistance and advice.

J.M.P. thanks Michael Brown and the Division of Geological and Planetary Sciences of the California Institute of Technology for sabbatical hospitality during the preparation of this paper.

Our work was funded in part by NASA grants NNG04GF99G and NNG04GK44G from the Solar Terrestrial Program and grant NNM07AA01G from NASA's Marshall Space Flight Center. A.C.S. was supported by funding from NASA's Office of Space Science through the Living with a Star, the Solar Physics Supporting Research and Technology, and the Sun-Earth Connection Guest Investigator Programs.